\documentclass[times]{article}
\usepackage{amsmath, amssymb, enumerate}

\usepackage{fancybox}
\usepackage{amsfonts}
\usepackage[usenames]{color}
\usepackage{listings}

\usepackage[linesnumbered,vlined]{algorithm2e}
\usepackage{graphicx}
\usepackage{times}
\usepackage{psfrag}
\usepackage{subfigure}
\usepackage{caption}
\usepackage{multirow}
\usepackage{epsfig}
\usepackage{url}

\usepackage{color}
\def\fixme#1{\typeout{FIXED in page \thepage : {#1}}
\bgroup \color{red}{} \egroup}

\usepackage{rotating,tabularx}

\makeatletter
\def\url@leostyle{%
\@ifundefined{selectfont}{\def\UrlFont{\sf}}{\def\UrlFont{\small\ttfamily}}}
\makeatother

\newcommand{\bottomrule}{\hline}
\newcommand{\toprule}{\hline}
\newcommand{\midrule}{\hline}
\begin{document}

\title{Parallelism-Aware Memory Interference Delay Analysis for COTS Multicore Systems}

\author{Heechul Yun \\
University of Kansas, USA. \\
heechul.yun@ku.edu\\
}

\maketitle
\thispagestyle{empty}
\begin{abstract}

In modern Commercial Off-The-Shelf (COTS) multicore systems, each core
can generate many parallel memory requests at a time. The processing of
these parallel requests in the DRAM controller greatly affects the memory
interference delay experienced by running tasks on the platform.

In this paper, we model a modern COTS multicore system which has a
non-blocking last-level cache (LLC) and a DRAM controller that
prioritizes reads over writes. To minimize interference, we focus on
LLC and DRAM bank partitioned systems. Based on the
model, we propose an analysis that computes a safe upper bound for
the worst-case memory interference delay.

We validated our analysis on a real COTS multicore platform with a set
of carefully designed synthetic benchmarks as well as SPEC2006
benchmarks. Evaluation results show that our analysis is more
accurately capture the worst-case memory interference delay and provides
safer upper bounds compared to a recently proposed analysis which
significantly under-estimate the delay.

\end{abstract}

\section{Introduction}
%
In modern Commercial Off-The-Shelf (COTS) multicore systems, many
parallel memory requests can be
sent to the main memory system at any given time for the following two
reasons. First, each core employs a variety of techniques---such as
non-blocking cache, out-of-order issues, and speculative
execution---to hide memory access latency. These techniques allow the
core continue to execute new
instructions while it is still waiting memory requests for previous
instructions to be completed. Second, multiple cores can run multiple
threads, each of which generates memory requests.

These parallel memory requests from the processor put high pressure
on the main memory system. To deliver high performance, modern DRAM
consists of multiple resources called banks that can be accessed in
parallel. For example, a typical DRAM module has 16 banks, supporting
up to 16 parallel accesses~\cite{micronddr3}. To efficiently utilize
the available bank level parallelism,
modern COTS DRAM controllers employ sophisticated techniques such as
out-of-order request processing, overlapped request dispatches, and
interleaved bank mapping~\cite{rixner2000memory,natarajan2004study,chatterjee2012staged}.

While parallel processing of multiple memory requests generally
improves overall memory performance, it is very difficult to understand
precise memory performance especially when multiple applications run
concurrently, because each memory request is more likely to be
interfered by other requests. Therefore, reducing interference and
improving performance isolation in COTS multicore systems has been an
important research topic in the real-time systems community.

To this end, software based DRAM bank partitioning ~\cite{yun2014rtas,liu2012software,suzuki2013coordinated}
and last-level cache (LLC) partitioning ~\cite{zhang2009towards, mancuso2013rtas,
ding2011srm, ward2013rtas, lin2008gaining} have been studied by many
researchers. These approaches reduce interference by allocating
dedicated cache space and/or DRAM banks. While effective, it is also
shown that partitioning these resources alone does not provide ideal
isolation due to interference in other parts of the memory
hierarchy, most notably in the DRAM
controller and the shared memory bus (command and data) which connects
the controller and the DRAM module ~\cite{yun2014rtas,
kim2014rtas}. As a result, it is still difficult to understand
worst-case memory interference delay, even when the LLC and DRAM banks
are partitioned.

Recently, Kim et al. proposed an analysis method that takes the DRAM
controller and the shared memory bus into
account~\cite{kim2014rtas}. They faithfully model a modern COTS DRAM
system, and provide an analytic upper bound on the worst-case memory
interference delay of each memory request of the task under analysis.
However, their analysis made a significant assumption that makes it
difficult to apply it to modern COTS multicore systems. Specifically, the
analysis assumes that each core can only generate one outstanding
memory request at a time and stalls until the memory request is
served. Unfortunately, this assumption is far from reality in modern
COTS platforms. As outstanding requests can interfere with the
memory request under analysis, the actual worst-case depends on the
number of outstanding requests, which is typically substantially
higher than the core count. For example, a COTS multicore processor
used in our evaluation supports up to 32 outstanding reads and 16
outstanding writes while it has only 4 cores. As a result, the computed
memory interference bounds can be significantly optimistic than the
reality (i.e., underestimating the actual delay), as we experimentally demonstrated
in Section ~\ref{sec:result_syn}.

In this work, we present a parallelism-aware memory interference delay
analysis. We model a COTS DRAM controller that has a
separate read and a write request buffer. Multiple outstanding memory
requests can be queued in the buffers and processed in
out-of-order to maximize memory performance. Also, reads are
prioritized over writes in our model. These features are commonly
found in modern COTS multicore systems and crucially important in
understanding memory interference.
To minimize interference, we only consider a system in which the LLC and
DRAM banks are partitioned. This is easily achievable on COTS multicore
systems via software~\cite{liu2012software,suzuki2013coordinated}.
Based on the system model, our analysis provides a safe analytic upper
bound on the worst-case memory interference delay for each memory
request of the task under analysis.

We evaluate the proposed analysis on a real COTS multicore platform
with a set of synthetic benchmarks as well as
SPEC2006 benchmarks. The synthetic benchmarks are specially
designed to simulate worst possible memory interference delay. As for
synthetic benchmarks, our analysis provides a tight and safe upper
bound while the analysis in ~\cite{kim2014rtas} significantly
under-estimates the actual delay (almost double than the computed
delay). As for SPEC2006 benchmarks, we
found our analysis provides safe upper bounds for all but
two benchmarks, while the compared analysis under-estimates 11 (out of 19)
benchmarks. Investigating the two benchmarks that our analysis
under-estimated (so did ~\cite{kim2014rtas}),
we find that space competition in miss status holding registers
(MSHRs)~\cite{kroft1981lockup}, which track the status of outstanding
cache misses, can be a considerable source of additional interference
in modern COTS systems.

Based on our analysis and empirical evaluation, we propose two simple
architectural supports, which can be easily
incorporated in modern COTS, to effectively reduce worst-case
interference delay.


Our contributions are as follows:
\begin{itemize}
\item To the best of our knowledge, our work is the first that considers
memory level parallelism and read prioritized DRAM controllers to
analyze memory interference delay.
\item We experimentally validate and compare our analysis with a state
of art analysis on a real COTS multicore
platform with a set of carefully designed synthetic benchmarks as well
as SPEC2006 benchmarks.
\item We propose two simple architectural supports that can
significantly reduce worst-case memory interference delay on COTS multicore
processors.
\end{itemize}

The remaining sections are organized as follows:
Section~\ref{sec:background} provides background on COTS multicore
systems and LLC and DRAM bank partitioning techniques.
Section~\ref{sec:motivation} discusses the state-of-art memory
interference delay analysis.
We present our analysis in Section~\ref{sec:analysis} and provide
evaluation results in Section~\ref{sec:evaluation}.
Section~\ref{sec:recommendation} discusses architectural recommendations.
Section \ref{sec:related} discusses related work.
Finally, we conclude in Section~\ref{sec:conclusion}.
\section{Background: Modern COTS Multicore systems} \label{sec:background}

\begin{figure} [t]
\centering
\centering
\includegraphics[width=0.40\textwidth]{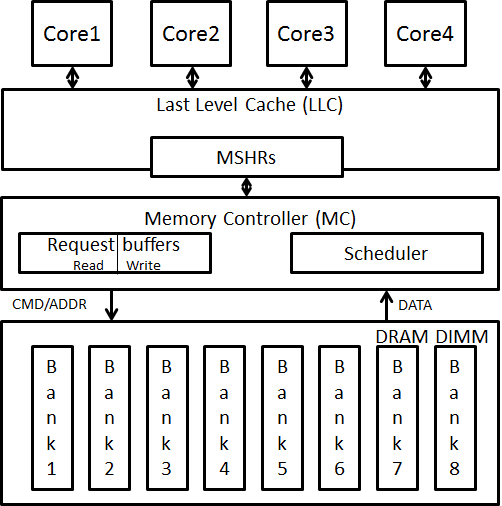}
\caption{Modern COTS multicore architecture}
\label{fig:architecture}
\end{figure}

A modern COTS multicore system, shown in
Figure~\ref{fig:architecture}, supports a high degree of memory
level parallelism through a variety of architectural features.
In this section, we provide some background on important
architectural features of modern COTS multicore systems, and review
existing software based resource partitioning techniques.

\subsection{Non-blocking Cache and MSHR}
At the cache level, non-blocking caches are used to handle multiple
simultaneous cache-misses. This is especially crucial for the shared last
level cache (LLC), as it is shared by all cores. The state of the
outstanding memory requests are maintained by a set of miss status
holding registers (MSHRs). On a cache-miss, the LLC allocates a
MSHR entry to track the status of the ongoing request and the entry is
cleared when the corresponding memory request is serviced from the
main memory. As such, the number of MSHRs effectively determines the
maximum number of outstanding memory requests directed to the DRAM
controller.

\subsection{DRAM Controller}
The DRAM controller receives requests from the LLC (or other DMA
devices) and generates DRAM specific commands to access data in the
DRAM. Modern DRAM controllers often include separate
read and write \emph{request buffers} and \emph{prioritize reads} over
writes because writes are not on the critical path for
program execution. Write requests are buffered on the write buffer of
the DRAM controller and serviced when there are no pending read
requests or the write queue is near
full~\cite{natarajan2004study,chatterjee2012staged}.
The DRAM controller and the DRAM module are connected
through a command/address bus and a data bus. Modern DRAM modules are organized
into ranks and each rank is divided into multiple
\textit{banks}, which can be accessed in \emph{parallel} provided that
no collisions occur on either buses.
Each bank comprises a row-buffer and an array of storage
cells organized as \textit{rows} and \textit{columns}. In order to
access the data stored in a DRAM row, an activate
command (\textit{ACT}) must be issued to load the data into the row
buffer first before it can be read or written. Once the data is in the
row buffer, any numbers of subsequent read or write
commands (\textit{RD, WR}) can be issued to access data in the row. If, however, a
request wishes to access a different row from the same bank, the row
buffer must be written back to the array with a pre-charge
command (\textit{PRE}) first before the second row can be
activated.

\subsection{Memory Scheduling Algorithm}
Due to hardware limitations, the memory device takes time to perform
different operations and therefore timing constraints between various
commands must be satisfied by the controller. The operation and timing
constraints of memory devices are defined by the JEDEC standard
\cite{jedec}. The key facts concerning timing constraints are:
1) the latency for accessing a closed row is much longer than
accessing a row that is already open; 2) different banks can be
operated in parallel since there are no long timing constraints
between banks.
To maximize memory performance, modern DRAM controllers typically use
a first-ready first-come-first-serve
(FR-FCFS)~\cite{rixner2000memory} scheduling algorithm that
prioritizes:
\begin{enumerate}
\item Ready commands over non-ready commands,
\item Column (CAS) commands over row (RAS) commands,
\item Older commands over younger commands.
\end{enumerate}
This means that the algorithm can process memory requests in out-of-order of their
arrival times. Note that a DRAM command is said to be~\emph{ready}
when it can be scheduled immediately as it satisfies all timing
constraints imposed by previously scheduled commands and the current
DRAM status.

\subsection{DRAM Bank and Cache Partitioning}
In order to maximize memory level parallelism, most COTS DRAM
controllers employ a version of \emph{interleaved bank} addressing
strategy. Under this scheme, consecutive memory blocks in physical
address space, typically of the size of a memory page, are allocated
to different banks. This makes pending memory requests in the DRAM
controller are likely to target different banks, thereby maximizing
memory level parallelism. In the worst case, however, it is possible
that all programs allocate memory on the same bank, resulting in
much increased memory latency compared to the average case. This
dependency on run-time decisions by the memory allocator can be a
significant potential source of unpredictability.
Furthermore, since banks are interleaved, any core in the system can
access any bank. If two applications running in parallel on different cores
access two different rows in the same bank, they can force the memory
controller to continuously pre-charge the row buffer and open a new
row every time an access is performed. This loss of row locality can
result in a much degraded row hit ratio and thus a corresponding
latency increases for both applications.

Software bank partitioning~\cite{yun2014rtas,liu2012software,suzuki2013coordinated}
can be used to avoid the problems of shared banks. The technique
leverages the page-based virtual memory system of modern operating
systems and allow us to allocate memory to specific DRAM
banks. Each core, then, can be assigned to use its
private DRAM banks, effectively eliminates bank sharing among cores
without requiring any hardware modification.
Similar techniques can also be applied to partitioning the shared
LLC as explored in ~\cite{zhang2009towards, mancuso2013rtas,
ding2011srm, ward2013rtas, lin2008gaining}. It is shown that
partitioning DRAM banks and LLC substantially reduce memory
interference among the cores~\cite{yun2014rtas}.

However, the LLC cache space and DRAM banks are not the only shared
resources contributing to memory interference. Most notably, at the
DRAM chip level, all DRAM banks fundamentally share the common command
and data bus. Hence, contention in the buses can become a bottleneck.
Furthermore, as many memory requests can be buffered inside the DRAM
controller's request buffers, its scheduling policy can greatly affect
memory interference delay. Finally, at the LLC level, the MSHRs for
the LLC are also shared by all cores even if the cache space is
partitioned. We will show its performance impact in Section
~\ref{sec:result_spec}.

\textbf{Goal:} The goal of this paper is to analyze the worst-case
memory interference delay in a cache and DRAM bank partitioned
system, focusing mainly on delay in the DRAM controller and command
and data bus between the controller and the DRAM module.


\section{The State of Art Delay Analysis and the Problem}\label{sec:motivation}

\begin{figure*} [t]
\centering
\centering
\subfigure[Initial bank queue status] {
\includegraphics[width=0.14\textwidth]{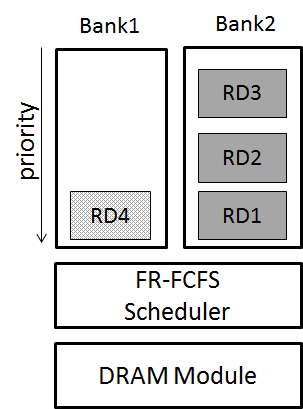}
}
\subfigure[Timing diagram under FR-FCFS schedule]{
\includegraphics[width=0.82\textwidth]{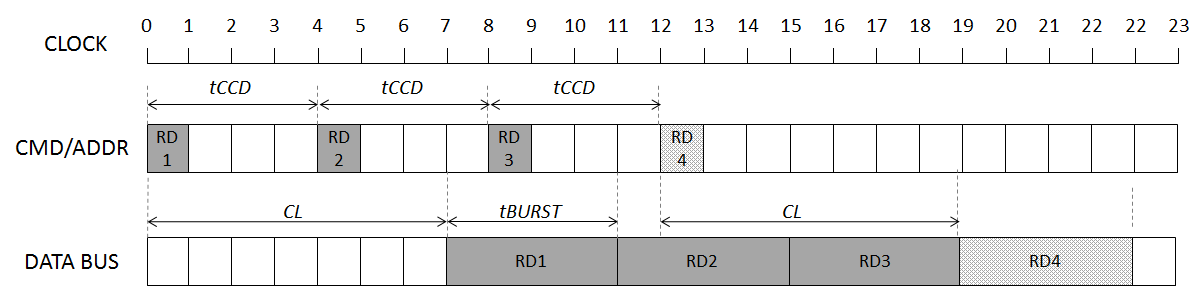}
}
\caption{Inter-bank delay caused by three previously arrived
outstanding requests. (DRAM commands are numbered according to
their arrival time to the DRAM controller.) }
\label{fig:motivation1}
\end{figure*}

\begin{figure*} [t]
\centering
\centering
\subfigure[Initial bank queue status] {
\includegraphics[width=0.14\textwidth]{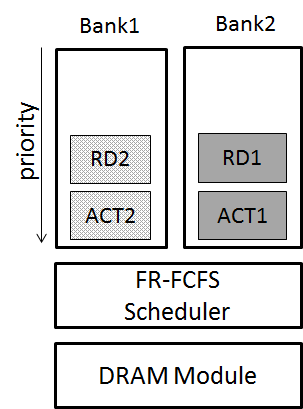}
}
\subfigure[Timing diagram under FR-FCFS schedule]{
\includegraphics[width=0.82\textwidth]{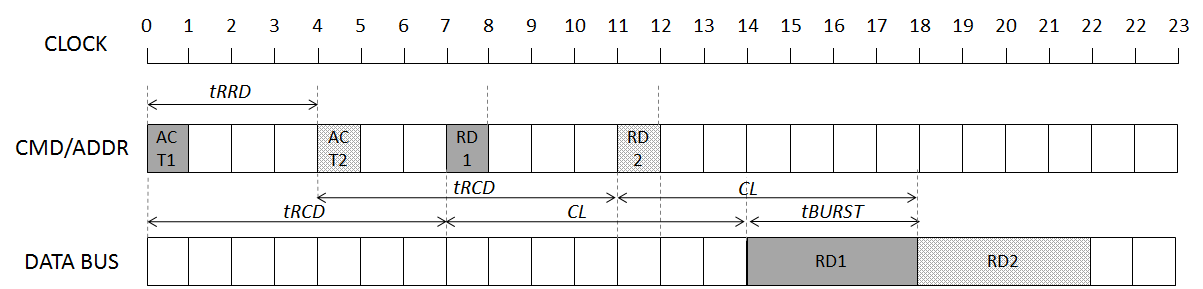}
}
\caption{Inter-bank delay caused by a previously arrived row-miss request.}
\label{fig:motivation2}
\end{figure*}

In this section, we first review a state of art memory interference delay
analysis for COTS memory systems, proposed by Kim et
al.~\cite{kim2014rtas}. We then investigate some of its assumptions
that are not generally applicable in modern COTS multicore systems.

The analysis models a modern COTS memory system in great detail. While
there has been a similar effort in the past ~\cite{zheng2013worst}, this
is the first work that considers the DRAM bank level request reordering
effect (i.e., out-of-order execution of young row-hit column requests
over older row-miss requests). Here, we briefly summarize the assumed
system model and part of the memory interference delay analysis,
relevant for the purpose of this paper.

The system model assumes a single channel DRAM controller and a DDR3
memory module. The DRAM controller
includes request buffers and uses the FR-FCFS scheduling algorithm.

At the high level, the analysis computes the worst-case memory
interference delay of the task under analysis $\tau_i$ either (1) as a
function of number of memory requests $H_i$ of the task (referred as
Request driven approach) or (2) as a function of the number of memory
requests generated by the other tasks on different cores (referred as
Job driven approach)---it takes the minimum of the two---similar to prior
work~\cite{yun2012ecrts}. The unique characteristics of the analysis
is that it considers both inter-bank and intra-bank (including request
reordering) memory interference delay. For the purpose of this paper,
however, we focus on their inter-bank delay analysis that assumes each
core is assigned \emph{dedicated DRAM bank partitions} using software bank
partitioning systems~\cite{yun2014rtas,liu2012software}.

The analysis assumes that each memory request of $\tau_i$ is composed
of PRE, ACT, and RD/WR DRAM commands (i.e., a row-miss) and each of
the command can be delayed by DRAM commands generated by other tasks
on different cores, due to inter-bank timing constraints imposed by
the JEDEC DDR3 specification~\cite{jedec}. These timing constraint
imposed inter-bank delay for PRE, ACT, and RD/WR commands are denoted
as $L_{inter}^{PRE}$, $L_{inter}^{ACT}$, and $L_{inter}^{RW}$,
respectively.

One major assumption of the analysis is that each core can generate
only \emph{one outstanding memory request} to the DRAM controller.
Based on this assumption, the worst-case per-request inter-bank memory
interference delay on a core $p$, $RD_p^{inter}$, is simply expressed by
$
RD_p^{inter} = \sum_{\forall q: q \neq p} \times ( L_{inter}^{PRE} + L_{inter}^{ACT} + L_{inter}^{RW}).
$

Finally, the total memory interference delay of a task is calculated
by multiplying $RD_p^{inter}$ to the number of total LLC misses $H_i$
of $\tau_i$

The analysis, however, has two main problems when it is applied to
modern COTS multicore systems. On the one hand, it is
overly \emph{optimistic} as it assumes each interfering core only can
generate one outstanding memory request at a time. Hence, it
essentially limits the maximum number of competing requests to the
number of cores in the system. However this is far from reality as
modern COTS multicore can generate many parallel memory requests at a
time. For example, a quad-core processor used in our evaluation can
generate up to 48 concurrent DRAM requests at a time (see
Section~\ref{sec:setup} for details). Because DRAM
performance is much slower than CPU performance, these requests are
queued inside the DRAM controller and can aggravate the overall delay.
Figure~\ref{fig:motivation1} illustrates
this problem. In the figure, three parallel requests RD1, RD2, and RD3
are already in the command queue for Bank2, when the request RD4 has
arrived at Bank1. Note that the DRAM commands are numbered in the
order of their arrival times in the DRAM controller. At memory clock
0, both RD1 and RD4 are ready, but RD1 is scheduled as FR-FCFS policy
prioritizes older requests over younger ones. Similarly, RD2 and RD3
are prioritized over RD4 at time 4 and 8, respectively. At other times
such as at clock 1, RD4 cannot be scheduled due a channel timing
constraint ($tCCD$), even though it
is ready w.r.t. the Bank1.

On the other hand, it is also overly \emph{pessimistic} as a memory
request---composed of PRE, ACT, and RD/WR DRAM sub-commands---is
assumed to suffer inter-bank interference for each sub-command, while
in reality the delays of executing sub-commands of a memory request
are not additive on efficient modern COTS memory
controllers.
Figure~\ref{fig:motivation2} shows such a case. In the
figure, each bank has one row miss DRAM request. Hence, each has to
open a new row with a ACT command followed by a RD command. Following
the FC-FRFS policy, ACT1 on Bank2 is executed first at clock
0. Even though ACT2 is targeting to a different bank, it is not
scheduled immediately due to the required minimum separation time $tRRD$
between two inter-bank ACT commands. At clock 4, however, ACT2 can be
issued even though ACT1 on the Bank2 is still in progress. In other
words, the two memory requests are \emph{overlapped}. Hence, when RD2
is finally issued at time 11, there is no extra inter-bank delay other
than the initial delay of $tRRD$.

From the point of view of WCET analysis, the former problem is more
serious as it undermines the safety of the computed WCET.

We experimentally validated the former problem on our test platform with
carefully engineered synthetic tasks, as we will detail in
Section~\ref{sec:result_syn}. To summarize the result, the calculated
worst-case response time using the stated analysis is up to 53\% smaller
than the measured worst-case response time. The result motivates our
analysis in the next section.
\section{Parallelism-Aware Memory Interference Delay Analysis} \label{sec:analysis}

In this section, we present our parallelism-aware memory interference
delay analysis that is aimed to support modern COTS multicore
systems. We begin by defining the system model on which our analysis
is based. We then present the main analysis with examples.

\subsection{System Model}

We consider a modern multicore architecture described in
Section~\ref{sec:background}. Specifically, there are $N_{proc}$ identical cores
in a single processor chip. A single LLC and MSHRs are shared among
the cores. When there is a miss in the LLC, an entry is registered on
the MSHR and it is removed when the associated DRAM transaction is
completed.
We assume a typical shared L3 cache that employs write-back
write-allocate policy. Hence
a write to DRAM only occurs when there is a L3 miss (either read or
write) that evicts a modified cache-line in the L3 cache, and
program execution can proceed without waiting the write request to be
processed in the DRAM. Therefore, for the analysis purpose, we only
consider memory interference delay imposed to each read request of the
task under analysis $\tau_i$. Note that the number of DRAM read requests
$H_i$ is equal to the number of LLC misses because, in a write-back
write-allocate cache, a write miss also generates a DRAM read request to
allocate the line in the L3 cache and then write to it.

On the DRAM controller side, we assume a modern DRAM controller that
supports the FR-FCFS scheduling
policy~\cite{rixner2000memory,wang2005umd} which is connected to a DDR3
DRAM module. At each memory clock tick, we assume a highly efficient
FR-FCFS scheduler that picks the highest priority ready command among all
requests and can overlap multiple requests simultaneously as long as
DRAM bank and channel level timing constraints and the
FR-FCFS priority rules are satisfied~\cite{natarajan2004study}.
We also assume that the DRAM controller has a read request buffer and a
write request buffer, and prioritizes reads over writes. The writes are
only serviced when there is no pending read request or the
write buffer is full.
The maximum number of prior read requests queued in the read request
buffer is denoted as $N_{rq}$ and we assume it is much bigger than
$N_{proc}$. In processing write requests, we assume it processes at
least $N_{wq}$ requests in a batch to amortize the cost of the bus
turnaround delay ~\cite{chatterjee2012staged}.
The values of $N_{rq}$ and $N_{wq}$ are platform specific and
determined by the number of MSHRs, the size of read request buffer and
the write-scheduling algorithm in the DRAM controller.
Table~\ref{tbl:sysparams} shows the parameters we used throughput this
paper which closely model our evaluation platform described in
Section~\ref{sec:setup}. ~\footnote{
Each core in our evaluation platform can have up to 10 outstanding
memory requests~\cite{david2012bandit}. Hence, $N_{rq} = 10 \times
(N_{proc} - 1) = 30$. As for $N_{wq}$, we use the value of Intel 870
memory controller ~\cite{natarajan2004study}.}

All previously mentioned assumptions closely follow common
behaviors of commercial COTS DRAM controllers~\cite{natarajan2004study}.
We assume open-page policy is used for bank management to
maximize data locality.
We assume a single rank DRAM module for simplicity but our analysis
can be extended to consider a multi-rank DRAM module.

We assume DRAM banks and the LLC space are partitioned on
a per-core basis. In other words, each core is assigned its own private
DRAM banks and LLC space. This can be easily achieved by using
software partitioning techniques on COTS
systems~\cite{yun2014rtas,liu2012software}.

Finally, we assume that any increase in memory latency is additive to
the task's execution time as in~\cite{kim2014rtas}. This is a
pessimistic assumption given that we consider out-of-order cores that
can hide much of memory access latency. Modeling reduced memory
latencies by OoO cores is, however, out of the scope of this paper.

In short, our system model is similar to ~\cite{kim2014rtas}, but
significantly differs in that (1) it models multiple parallel memory
requests buffered in the DRAM controller, and (2) it maintains
separate read and write request queues in the DRAM controller and
reads are prioritized over writes. Both are common characteristics
of modern COTS multicore memory systems ~\cite{natarajan2004study}.

Lastly, Table~\ref{tbl:dramparams} shows the DRAM parameters we used
throughout this paper.








\begin{table}[htbp]
\centering
\caption{System parameters for our evaluation platform}
\begin{tabular}{rrr}
\toprule
Symbols & Description & Value \\
\midrule
$N_{rq}$ & Maximum no. of prior read requests & 30 \\
$N_{wq}$ & Maximum no. of prior write requests & 4 \\
$N_{proc}$ & Number of cores & 4 \\
\bottomrule
\end{tabular}%
\label{tbl:sysparams}%
\end{table}%

\begin{table}[htbp]
\centering
\caption{DRAM timing parameters~\cite{micronddr3}}
\begin{tabular}{rrrr}
\toprule
Symbols & Description & DDR3-1066 & Units \\
\midrule
$tCK$ & DRAM clock cycle time & 1.87 & nsec \\
$tRP$ & Row precharge time & 7 & cycles \\
$tRCD$ & Row activation time & 7 & cycles \\
$CL$ & Read latency & 7 & cycles \\
$WL$ & Write latency & 6 & cycles \\
$tBURST$& Data burst duration & 4 & cycles \\
$tCCD$ & Column-to-Column delay & 4 & cycles \\
$tWTR$ & Write to read delay & 4 & cycles \\
$tRRD$ & Activate to activate delay & 4 & cycles \\
$tRTP$ & Read to precharge delay & 4 & cycles \\
$tFAW$ & Four activate windows & 20 & cycles \\
$tRC$ & Row cycle time & 27 & cycles \\
\bottomrule
\end{tabular}%
\label{tbl:dramparams}%
\end{table}%

\begin{figure*} [t]
\centering
\centering
\subfigure[Initial bank queue status] {
\includegraphics[width=0.14\textwidth]{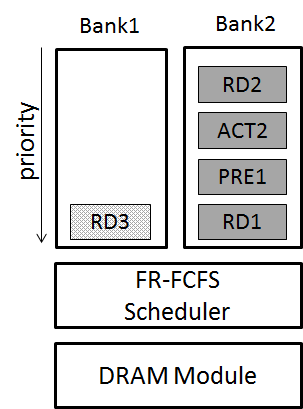}
}
\subfigure[Timing diagram under FR-FCFS schedule]{
\includegraphics[width=0.82\textwidth]{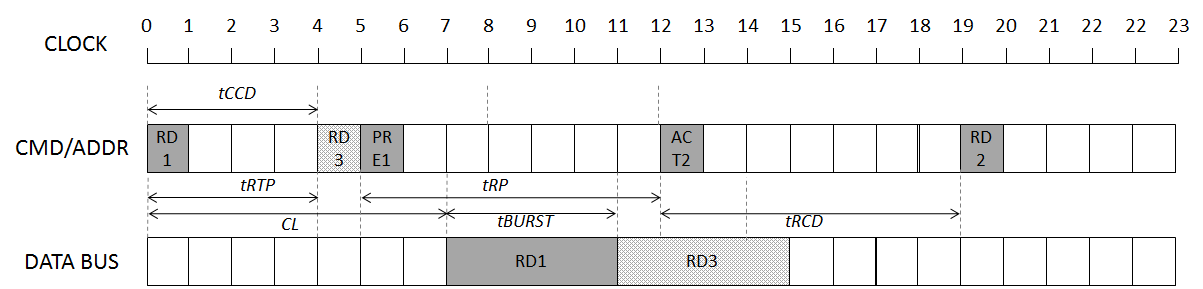}
}
\caption{Out-of-order request processing due to bank timing constraints.}
\label{fig:timing-reorder}
\end{figure*}

\begin{figure*} [t]
\centering
\centering
\includegraphics[width=0.95\textwidth]{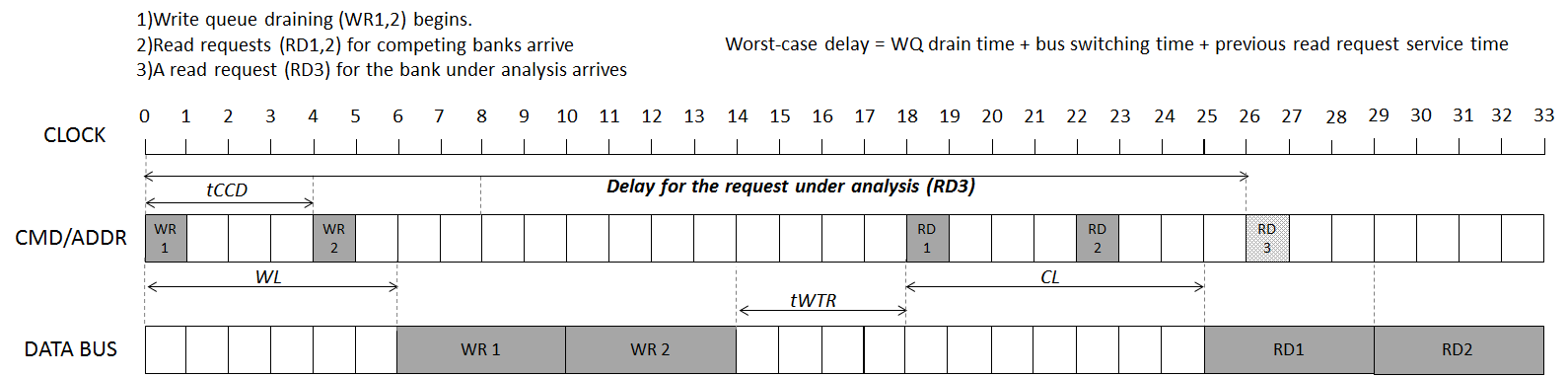}
\caption{Worst-case per-request inter-bank interference delay.}
\label{fig:timing-worst}
\end{figure*}

\subsection{Delay Analysis}

We now present our analysis that considers parallel memory requests
in modern COTS multicore systems.

As mentioned in the previous section, write memory requests are not in
the critical path of program execution in modern COTS systems. Hence,
our primary concern is memory interference delay to read
requests of the task under analysis.

As we consider a system in which both the LLC cache and DRAM banks
are partitioned on a per-core basis, conventional cache space and share bank
level contention do not exist. However, because the command and data
bus are shared in processing the queued memory requests, DRAM
controller's request scheduling greatly affects memory interference
delay. We now detail our delay analysis. Because the DRAM
controller prioritize reads and the bus turn-around cost is high, the
DRAM controller process requests in a \emph{batch} for either reads or
writes. In each processing mode, we analyze the worst-case memory
interference delay to a newly arrived read request.


\subsubsection{Data Bus Contention Delay}

When the DRAM controller is in the read processing mode, the
worst-case to a newly arrived read request occurs when the request
buffer is fully occupied by previously arrived $N_{rq}$ read requests
from the other competing cores. Furthermore, regardless whether the
read request under analysis is row-hit or row-miss, the worst-case
interference delay occurs when all the previous reads are pipelined
(i.e., overlapped scheduling~\cite{natarajan2004study}).

Reads can be pipelined in two cases: consecutive reads on the
same row or reads over different banks (see Figure~\ref{fig:motivation2}).
When reads are pipelined, the data bus is fully occupied and the
newly arrived read (the request under analysis) must wait until all
the previous reads are processed (because FR-FCFS prioritize older
requests). If, however, the previous requests are not
pipelined, the read request under analysis (younger request) can be
processed ahead of older requests on the other banks
(i.e. out-of-order processing). Figure~\ref{fig:timing-reorder} shows
such an example. At time 0, both RD3 and RD1 are ready to be scheduled
and FR-FCFS schedules RD1 as it is older than RD3. At time 4, both
PRE1 and RD3 are ready, but this time FR-FCFS chooses RD3 as it first
prioritizes row-hit column commands (i.e., RD) over other commands
(i.e., PRE, ACT).

When read requests are pipelined, the processing time of each read is
$tBURST$. Therefore, the delay caused by previously arrived read
commands $L_{rq}$ is
\begin{equation}
L_{rq} = N_{rq} \times tBURST.
\end{equation}

Note that if a read request under analysis needs to execute PRE or ACT
commands (closing the previous row and open a new row, respectively),
they can be executed in parallel by the time the data bus becomes
free, without adding to the total delay.

\subsubsection{Write Draining and Bus Turn-around Delay} \label{sec:wqdrain}

When there is no pending reads or the write request buffer is full,
the DRAM controller switches the mode to process pending writes. It is
called write draining and once the drain process begins, new incoming
read requests must wait until at least
$N_{wq}$ writes are drained to amortize the bus switching cost
$tWTR$~\cite{chatterjee2012staged}.

In draining writes, the worst-case happens when all writes access
different rows in the same bank, forcing the memory
controller to close and open a new row for each write. In this case, the
required time between two successive writes is determined by the row
cycle time $tRC$.

Therefore, the write queue draining delay $L_{wq}$ is given by
\begin{equation}
\begin{split}
L_{wq} = N_{wq} \times tRC + tWTR. \\
\end{split}
\label{eq:wqdrain}
\end{equation}

Then, the worst-case delay for a read request for the core under analysis
arises when the request arrived right after (1) the write queue drain process
began and (2) $N_{rq}$ read requests from other cores
arrived. A simplified illustrative example is shown in
Figure~\ref{fig:timing-worst}. In the figure, at time 0, three events
occurred in-order: (1) the write queue drain started to process two pending
write requests (WR1 and WR2); (2)two read requests (RD1 and RD2) from
competing cores arrived; (3) a read request (RD3) from the core under
analysis arrived. In this case, the RD3 must wait until all previous
activities finish.

Therefore, the worst-case inter-bank delay $D_p$ for a read request on
the core under analysis $p$ is expressed as follows:

\begin{equation}
D_{p} = L_{rq} + L_{wq}.
\end{equation}

Finally, the total inter-bank memory delay of $t_i$ can be computed by
\begin{equation}
H_i \times D_p,
\end{equation}
where $H_i$ is the number of LLC misses.
\section{Evaluation}\label{sec:evaluation}

In this section, we first present details on the hardware and software
platform used in our evaluation. We then present our evaluation
results obtained using a set of synthetic and SPEC2006 benchmarks.

\subsection{Evaluation Setup} \label{sec:setup}

Our hardware platform is a quad-core Intel Xeon W3530
(Nehalem)~\cite{molka2009memory} based computer. Each core has a
private L1 cache (32K-I/32K-D) and a private L2 cache (256~KiB),
and all cores share a 8MiB L3 cache. Shared MSHRs (called Global
Queue or GQ~\cite{intel2012optimization}) track the status of
up to 32 read requests and 16 write requests from all cores. According
to ~\cite{david2012bandit}, a single core can generate up to 10
concurrent read requests at a time, which we also experimentally
verified to be true in our test platform.
The memory controller (MC) is integrated in the processor and clocked
at 1066~MHz. The computer equips a single-channel
dual-rank 4~GiB PC10666 DDR3 DIMM module which includes 16 DRAM
banks. We disabled all hardware prefetchers, dynamic voltage
and frequency scaling, and the turbo-boost feature for better
predictability.

We use PALLOC~\cite{yun2014rtas} to partition DRAM banks and the L3
cache. For the purpose of our evaluation, we assign one private DRAM bank and
1/4 (2MiB) private L3 cache partition to each core. Therefore, there are
neither cache space evictions nor DRAM bank conflicts caused by
memory accesses from contending cores.

For measurement, we use Linux kernel's $perf$ infrastructure to
monitor LLC miss hardware performance counter.

\subsection{Results with Synthetic Benchmarks} \label{sec:result_syn}

\begin{figure} [t]
\centering
\centering
\includegraphics[width=0.45\textwidth]{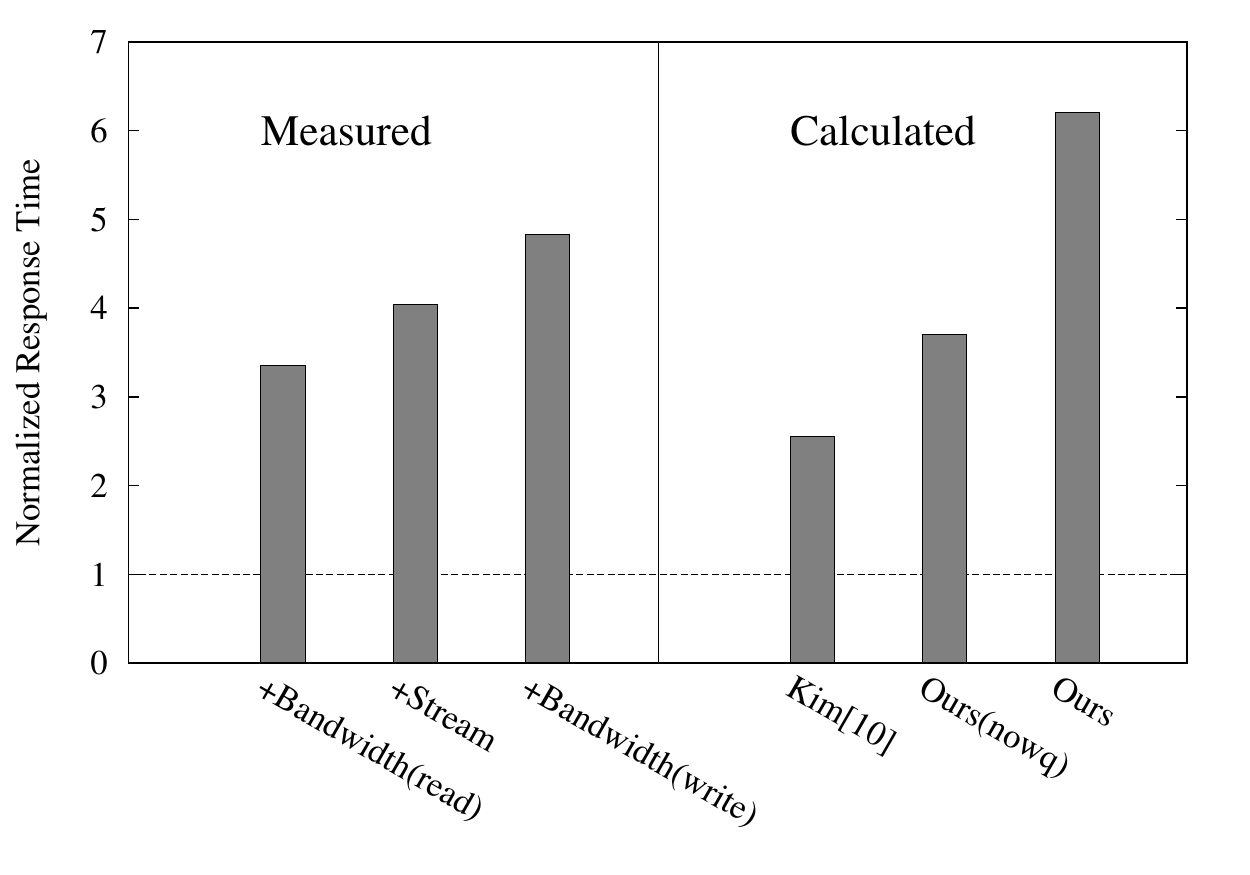}
\caption{Measured and analytic worst-case response times of
\emph{Latency} benchmark under high memory interference.}
\label{fig:result_syn}
\end{figure}

We now investigate the validity of our analysis compared to
experimental results obtained using a set of carefully engineered
synthetic benchmarks.

In this experiment, our goal is to simulate and measure the worst-case
memory interferences on a system in which DRAM banks and the LLC are
partitioned. We use \emph{Latency} benchmark~\cite{yun2013rtas} as
the task under analysis. The benchmark is a pointer-chasing application
over a randomly shuffled linked-list. Due to data dependency, it only
can generate one outstanding memory request at a time. Furthermore,
because the size of linked list is two times bigger than the size of
the LLC, each memory access is likely to result in a cache miss, hence
generating a DRAM request. As a result, its execution time highly
depends on DRAM performance and any delay in its memory access will
directly contribute to its execution time increase. In effect, this
benchmark defeats any potential benefits from out-of-order instruction
processing and other memory latency hiding techniques (i.e., an
equivalent of in-order processing).

We first run the Latency benchmark alone on Core0 to collect
its solo execution time and the number of LLC misses. We then
co-schedule three memory intensive tasks on the
other cores (Core1-3) to generate high memory traffic and measure the
response time increase of the Latency benchmark.
Note that the number of L3 misses of Latency do not change between
solo and co-scheduled execution as the L3 cache space is
partitioned. Furthermore, because each core also has a dedicated DRAM
bank, the number of DRAM row hit and misses also would not
change. Therefore, any response time increase mainly comes from
contention in the DRAM controller and its shared command and
data bus which we modeled in Section~\ref{sec:analysis}.
We repeat the experiment with three different memory intensive
benchmarks: \emph{Bandwidth(read)}, \emph{Bandwidth(write)}, and
\emph{Stream}\footnote{We use the code provided by the authors of
\cite{kim2014rtas}.}. All three benchmarks essentially access a big
array continuously but differ in their access
patterns---Bandwidth(read) only performs
consecutive reads; Bandwidth(write) do writes only; Stream performs both
reads and writes. Because memory accesses of these benchmarks do not
have data dependencies, modern Out-of-Order (OoO) cores can generate
as many outstanding requests as possible, hence simulating the
worst-case as their requests will occupy most of the read request
buffer (and the write request buffer) in the DRAM controller.

Figure~\ref{fig:result_syn} shows both measured and analytically
calculated response times of the Latency benchmark (normalized to the
solo execution time). First, measured response times in the
left side of the figure show that Bandwidth(write) causes the
highest memory interference. This is because the benchmark generates
two memory requests---a DRAM read (a cache-line allocation) and a DRAM
write (a write-back)---for each LLC miss, and processing writes can add
high delays due to reasons described in Section~\ref{sec:wqdrain}.
Second, note that the state-of-art
analysis~\cite{kim2014rtas} significantly under-estimates the memory
interference delay---the computed WCET is just 53\% of the measured
worst-case response time. This is mainly due to the fact that the
analysis assumes only one memory request from each competing core
while in this experiment competing cores generate many requests at a
time occupying the request buffers. On the other hand, our analysis,
denoted as \emph{Ours}, provides a safe upper bound for all
cases. Note that our analysis that ignores write-queue induced
worst-case latency, denoted as \emph{Ours(nowq)}, provides an upper
bound when the co-scheduled task
performs read only---i.e., Bandwidth(read)---but fails to do so when
the co-scheduled task performs many writes---i.e., Stream and
Bandwidth(write)---because it does not account additional
delay caused by occasional write buffer draining.
Lastly, the calculated WCET of our analysis is considerably
higher than the measured worst-case response
time by about 29\%. We believe this is because
our analysis assumes that all
writes are row-misses (see Eq.~\ref{eq:wqdrain}) in draining the
write-queue, while the actual writes from the benchmark are mostly
row-hits. In other words, the analysis over-estimated write-queue
draining delay $L_{wq}$.

\begin{table}[t]
\centering
\caption{SPEC2006 benchmark characteristics}
\begin{tabular}{rrrc}
\toprule
\multirow{2}[0]{*}{Benchmark} & Average & LLC misses & Memory \\
& IPC & per msec & intensity \\
\midrule
462.libquantum & 0.52 & 32497 & \\
482.sphinx3 & 0.70 & 22429 & \\
437.leslie3d & 0.39 & 21478 & high \\
450.soplex & 0.24 & 17970 & \\
471.omnetpp & 0.30 & 16629 & \\
\hline
403.gcc & 0.89 & 8465 & \\
483.xalancbmk & 0.11 & 7035 & \\
465.tonto & 1.21 & 5995 & \\
447.dealII & 1.41 & 4941 & medium \\
445.gobmk & 1.12 & 2531 & \\
456.hmmer & 1.94 & 2001 & \\
454.calculix & 2.38 & 1970 & \\
458.sjeng & 1.33 & 1672 & \\
435.gromacs & 1.12 & 1334 & \\
\hline
400.perlbench & 0.37 & 907 & \\
464.h264ref & 1.92 & 759 & \\
444.namd & 1.61 & 372 & low \\
416.gamess & 2.08 & 40 & \\
453.povray & 1.35 & 0 & \\
\bottomrule
\end{tabular}%
\label{tbl:spec2006}%
\end{table}%

\begin{table}[htbp]
\centering
\caption{Normalized response times of SPEC2006 benchmarks
with three memory intensive tasks.}
\begin{tabular}{r|c|rr||rr}
\toprule
\multirow{2}[0]{*}{Benchmark} & \multirow{2}[0]{*}{Measured} &
\multicolumn{2}{c||}{Calculated} &
\multicolumn{2}{c}{Pessimism} \\
& & Kim\cite{kim2014rtas} & Ours
& Kim\cite{kim2014rtas} & Ours \\
\midrule

462.libquantum & 3.22 & 5.19 & 15.10 & 61\% & 369\% \\
482.sphinx3 & 3.31 & 3.89 & 10.73 & 18\% & 224\% \\
437.leslie3d & 2.45 & 3.77 & 10.32 & 54\% & 321\% \\
450.soplex & 2.45 & 3.32 & 8.80 & 35\% & 259\% \\
471.omnetpp & 3.01 & 3.15 & 8.21 & 4\% & 173\% \\
\hline
403.gcc & 2.53 & 2.09 & 4.66 & -17\% & 85\% \\
483.xalancbmk & 1.68 & 1.91 & 4.05 & 14\% & 141\% \\
465.tonto & 1.78 & 1.77 & 3.60 & 0\% & 103\% \\
447.dealII & 1.59 & 1.64 & 3.14 & 3\% & 98\% \\
445.gobmk & 1.34 & 1.33 & 2.10 & -1\% & 57\% \\
456.hmmer & 1.32 & 1.26 & 1.87 & -5\% & 42\% \\
454.calculix & 1.31 & 1.25 & 1.85 & -4\% & 42\% \\
458.sjeng & 1.35 & 1.22 & 1.73 & -10\% & 28\% \\
435.gromacs & 1.20 & 1.17 & 1.58 & -2\% & 32\% \\
\hline
400.perlbench & 1.23 & 1.12 & 1.39 & -9\% & 14\% \\
464.h264ref & 1.18 & 1.10 & 1.33 & -7\% & 12\% \\
444.namd & 1.08 & 1.05 & 1.16 & -3\% & 7\% \\
416.gamess & 1.07 & 1.01 & 1.02 & -6\% & -5\% \\
453.povray & 1.35 & 1.00 & 1.00 & -26\% & -26\% \\

\bottomrule
\end{tabular}%
\label{tbl:result_spec}%
\end{table}%

\subsection{Results with SPEC2006 Benchmarks} \label{sec:result_spec}

In this subsection, we evaluate the response times of SPEC2006
benchmarks. The main characteristics of 19 benchmarks we
used are given in Table~\ref{tbl:spec2006}. We exclude 10 (out of 29)
benchmarks whose memory footprints are bigger than a DRAM bank
partition size (i.e., 256MB) for the purpose of our evaluation.

The basic experiment setup is the same as the previous subsection except
that we now use each of SPEC2006 benchmark as the task under analysis
instead of the Latency benchmark. As for interfering tasks, we use
Bandwidth(write) benchmark, described in the previous subsection, as
it gives worst-case memory interference.

Table~\ref{tbl:result_spec} shows the measured and analytic response
times. The two rightmost columns show pessimism in the analysis
compared to the measured response times. Note first that the baseline
analysis ~\cite{kim2014rtas} under-estimates worst-case response times
of 11 out of 19 benchmarks, although the degree of under-estimation is
much less than the engineered synthetic tasks we used in
Section~\ref{sec:result_syn}. As explained earlier, this is because
the analysis does not take multiple outstanding memory requests into
account, resulting much less queuing delay in its calculation than
reality.

Interestingly, both analyses
under-estimate the response times of 453.povray and
416.gamess. This is because both benchmarks generate very little
(close to zero) DRAM traffic, as can be see in
Table~\ref{tbl:spec2006}, the added interference is not caused by DRAM
related interference that both analyses try to estimate. This is
interesting as we already partition the L3 cache space among
cores. It means that the observed interference delay is caused neither by
cache space competition nor DRAM related interference.
To further investigate the source of the delay, we varied the number of
interfering tasks---i.e., Bandwidth(write)---from 1 to 3, and found
that performance suffers only when there are more than two Bandwidth
instances. We believe this is because the MSHRs are shared by both L2
and L3 caches in our platform so that cache misses of both caches
compete the limited MSHR space. Specifically, there are a total of 32
entries for outstanding reads where each core can use up to 10
entries. When three Bandwidth instances run (on Core1-3), they use up
to 30 entries (10 entries/core x 3 cores), it leaves only two
entries for the task under analysis (on Core0)---\emph{a 80\% reduction}
(2 out of 10). Given that both benchmarks (453.povray and 416.gamess)
show relatively high L2 miss rates, they likely suffer from the
reduction in the available MSHR entries.
We currently do not consider this MSHR space
contention in the analysis as we have no control over the allocations
of MSHRs. We will discuss how we can provide better isolation
concerning MSHR competition in Section~\ref{sec:recommendation}.

Other than these two benchmarks, our analysis provides safe upper
bounds for the rest of benchmarks we tested, albeit pessimistic. We
argue, however, this is expected behavior
given the fact that our analysis---as well as ~\cite{kim2014rtas}---does
not consider latency hiding techniques used in modern OoO cores, which
are highly effective in reducing perceived memory access latency to
the task~\cite{wang2002memory}, and we assume increased memory
access latency is addictive to the task execution time. Another major source
of pessimism in our analysis comes from the fact that we assume the
read request queue in the DRAM controller is always fully occupied by
prior requests from the interfering tasks. However, when the task under
analysis itself is highly memory intensive and has a high degree of memory level
parallelism, such as 462.libquantum, the read request queue likely
contains many memory requests from the analyzed task.



\section{Desired Architectural Support for Real-Time Systems} \label{sec:recommendation}
In this section, we discuss two simple and low-cost architectural supports
that can greatly reduce worst-case memory interference delay on COTS
multicore systems.

\subsection{Software Controlled MSHR Reservation}

MSHRs are important shared resources that determine the amount of parallelism in
the system. As experimentally shown in Section~\ref{sec:result_spec},
when a highly memory intensive task generate many parallel requests,
MSHRs become scarce, thereby significantly lower achievable
memory level parallelism of competing tasks. As a result, competing
tasks' performance would suffer. To achieve better performance
isolation, it is desirable for each core to reserve a fraction of
MSHRs, preferably by software. This can be easily implemented in
hardware, as shown in ~\cite{ebrahimi2010fairness}, and can eliminate
unintended memory interference due to contention in MSHRs.

\subsection{Software Controlled Bank Prioritization}

The biggest factor in high worst-case memory interference comes from
the fact that a large number of previously arrived memory requests are
prioritized under FR-FCFS scheduling policy, even if the newly arrived
request is from a higher priority task, effectively creating a
priority inversion problem~\cite{sha2004real}. Hence, from the
real-time perspective, it is highly desirable if software can
influence on prioritization logic of the DRAM controller. If, for
example, software can prioritize a specific bank over the other banks,
memory requests to the prioritized bank can always be processed
almost immediately without waiting the all the queued requests are
serviced. It will be especially effective for a DRAM bank partitioned
system as assumed in our analysis.
\section{Related Work} \label{sec:related}

As memory performance is becoming increasingly important in modern multicore
systems, there have been great interests in the real-time research
community to minimize and analyze memory related interference delay for
designing more predictable real-time systems.

Initially, many researchers model the cost to access the
main memory as a constant and view the main memory as a single resource
shared by the cores~\cite{yun2012ecrts,pellizzoni2010worst,yao2012memory,schranzhofer2010timing}. 
However, modern DRAM systems are composed of many sophisticated
components and the memory access cost is far from being a constant as it
varies significant depending on the states of the variety of components
comprising the system.

Many researchers turn to hardware approaches and develop specially
designed DRAM controllers that are highly predictable and provide
certain performance guarantees~\cite{reineke2011pret,zheng2013worst,akesson2007predator,paolieri2009analyzable,goossen2013conservative}.
The work in \cite{reineke2011pret} and \cite{zheng2013worst} both
implement hardware based private banking scheme which eliminate
interferences caused by sharing the banks. They differ in that the controller
in \cite{reineke2011pret} uses close page policy with TDMA scheduling
while the work in \cite{zheng2013worst} uses open page policy with
FCFS arbitration. AMC~\cite{paolieri2009analyzable} and Predator
\cite{akesson2007predator} utilize interleaved bank and close page
policy. Both approaches treat multiple memory banks as a single unit of
access to simplify resource management. They differ in that AMC uses a
round-robin arbiter while Predator uses the credit-controlled
static-priority (CCSP) arbitration \cite{akesson2008rtcsa}, which
assigns priorities to requestors in order to guarantee minimum bandwidth
and bounded latency. While these proposals are valuable,
especially for hard real-time systems, they are not available in COTS
systems.

To improve performance isolation in COTS systems, several recent papers
proposed software based bank partitioning
techniques~\cite{yun2014rtas,liu2012software,suzuki2013coordinated}. They
exploit the virtual memory of modern operating systems to allocate
memory on specific DRAM banks without requiring any other special
hardware support. Similar techniques has long been applied in
partitioning shared caches
~\cite{liedtke97ospart,lin2008gaining,zhang2009towards,soares2008reducing,ding2011srm,ward2013rtas,mancuso2013rtas}. These
resource partitioning techniques eliminate space contention of the
partitioned resources, hence improve performance isolation. However,
as shown in ~\cite{yun2014rtas,kim2014rtas}, modern COTS systems have
many other still shared components that affect memory performance.
A recent attempt to analyze these effects~\cite{kim2014rtas}, which is
reviewed in Section~\ref{sec:motivation}, greatly increased our
understanding on the DRAM controller, but its system model is still
far from real COTS systems, particularly on its assumption of one
outstanding memory request per core. In contrast, our work models
a more realistic COTS DRAM controller that handles multiple
outstanding memory requests from each core and out-of-order memory
request processing (i.e., prioritizing reads over writes). We believe
our system model and the analysis capture commonly found architectural
features in modern COTS systems, hence better applicable in analyzing
memory interference on COTS multicore systems.

\section{Conclusion} \label{sec:conclusion}
We presented a new parallelism-aware worst-case memory interference
delay analysis for COTS multicore systems. We model a COTS DRAM
controller that has a separate read and a write request
buffer. The modeled DRAM controller buffers multiple outstanding memory
requests from the LLC and processes them in out-of-order fashion. It
prioritizes reads over writes and row-hit over misses. By modeling
these architectural features, which are commonly found in COTS
multicore systems, our analysis can compute more accurate worst-case
memory access delay of COTS multicore systems.

We validated our analysis on a real COTS multicore platform with a set
of carefully designed synthetic benchmarks as well as SPEC2006
benchmarks. For synthetic benchmarks, our analysis produces a tight
and safe upper bound while the compared recent work~\cite{kim2014rtas}
significantly under-estimates the interference delay. For SPEC2006
benchmarks, our analysis is more pessimistic but safer than the
compared work. These evaluation results show that our analysis is
better applicable for modern COTS multicore systems.
As future work, we will examine several architectural supports that can
provide better isolation and reduce pessimism in the analysis.


\section*{Acknowledgements} \label{acknowledge}
This research is supported in part by NSF CNS 1302563.

\bibliographystyle{plain}
\bibliography{heechul}
\end{document}